\documentclass[%
 amsmath,amssymb,
 aps,
]{revtex4-2}

\usepackage{graphicx}
\usepackage{dcolumn}
\usepackage{bm}

\begin{document}

\title{Timely pandemic countermeasures reduce both health damage and economic loss: Generality of the exact solution}
		
\author{Tsuyoshi Hondou}
\email[]{hondou@mail.sci.tohoku.ac.jp}
\affiliation{Graduate School of Science, Tohoku University
\\ 
Sendai 980-8578, Japan}

\date{\today}

\begin{abstract} 
Balancing pandemic control and economics is challenging, as the numerical analysis assuming specific economic conditions complicates obtaining predictable general findings. In this study, we analytically demonstrate how adopting timely moderate measures helps reconcile medical effectiveness and economic impact, and explain it as a consequence of the general finding of ``economic irreversibility" by comparing it with thermodynamics. A general inequality provides the guiding principles on how such measures should be implemented. The methodology leading to the exact solution is a novel theoretical contribution to the econophysics literature.
\end{abstract}

\maketitle

Since early 2020, the world has been considerably impacted by the COVID-19 pandemic. To prevent widespread infections, several countries adopted severe countermeasures, including extended lockdowns, particularly when their medical care systems were on the verge of a breakdown. Therefore, how to adopt countermeasures against such a pandemic as well as future infections, has emerged as an important question because countermeasures such as lockdowns may significantly damage socioeconomic systems. As addressing this question needs economic analysis, mathematical economists began to examine this issue numerically \cite{CEPR,R&M,ERT,Acemoglu}. However, numerical findings assuming individual socioeconomic situations lack generality as they depend on specific parameters \cite{Saltelli}. 

To determine a general result, an analytic study that transcends conventional methodology and numerical methods of mathematical economics is indispensable. The history of physics leads us to acknowledge the importance of theory derived through analytic studies, in which we concentrate on the effect of a small number of elements essential to the system. The number of elements may be more in a real system than that used in theory. However, the exact solution of a simple model can provide deep insights and high predictability.

Given the physicist's viewpoint, the author was inspired by the pioneering numerical work of Rowthorn and Maciejowski \cite{R&M}, who used economic parameters for the United Kingdom to numerically demonstrate that keeping the infected population stationary after lockdown is economically optimum. The author then formulated a theoretical framework and found a preliminary analytic result, analogous to the thermodynamic theory, that the early adoption of a countermeasure, keeping the infected population constant, is more effective than a delayed countermeasure in reducing both infected population and economic damage \cite{JPSJ2021}. However, the result depends on the special condition that the infected population must be cyclic in time, in contrast to the realistic situation where the maximum number of infected people is important. Therefore, the theory cannot be applied to general situations that are out of the cycle.

In this study, the author formulates a theoretical framework and evaluates the countermeasure process for the pandemic outbreak (infection-spreading) phase \cite{note0} because it is the principal and the most important phase of a pandemic \cite{note0}. 
The early adoption of countermeasures to keep the exponential growth rate of the infected population moderate is shown to be better than delayed countermeasures implemented just before the crisis of medical care capacity in reducing both the infected population and economic damage. The result includes the preliminary findings \cite{JPSJ2021} under the cyclic condition as a special solution. It is also shown to be a natural consequence of the general finding of ``economic irreversibility." \cite{Ayres}

Pandemic control comprises measures adopted to reduce the number of people infected by an infected person. The average number within a society is called the ``effective reproduction number," $R_{\mbox{t}}$ \cite{BF}. When $R_{\mbox{t}}$ drops below 1, the pandemic subsides. Countermeasures, including wearing masks, strengthening ventilation, suspending business activities, and announcing lockdowns, can be adopted to reduce $R_{\mbox{t}}$ from its uncontrolled (natural) value, $R_{\mbox{N}} (>1) $. $R_{\mbox{N}} $ at the initial phase of infection equals the basic reproduction number $R_0$  \cite{BF}.
 
Hereafter, four general assumptions are adopted for the outbreak phase of the pandemic. 
\begin{enumerate}
\item[i)]
Infected population varies exponentially according to the effective reproduction number, $R_{\mbox{t}}$. 
\item[ii)]
 The social cost of countermeasures is defined by the intervention cost, where countermeasures are assumed to be adopted in sequential order of cost-effectiveness. To further decrease $R_{\mbox{t}}$, society must implement costlier measures \cite{R&M}.
\item[iii)]
 Medical costs, including costs triggered by the increase in the infected population, increase as the infected population increases. 
\item[iv)]
 The total cost is the sum of the intervention and medical costs. Note that the set of assumptions 2, 3, and 4 is a typical example of cost-benefit analysis in economics \cite{Boardman,Nas}. The current formulation regarding the cost-benefit analysis is similar to that in Rowthorn and Maciejowski \cite{R&M} and is conducted as follows.
\end{enumerate}

Assumption (i) yields the following equation for the infected population at time $t$, $I(t)$:
        \begin{equation}
        \frac{d I (t) }{d t}  = \gamma \Delta (t)  \, I(t) ,
            \label{dd}
        \end{equation}
where $\Delta (t) = R_{\mbox{t}} -1 $, and $\gamma$ is a constant that determines the time scale of the dynamics \cite{JPSJ2021}.
At $R_{\mbox{t}}=1$, the infected population is stationary as $\Delta (t)= 0$. As exponential dynamics is a general property in a pandemic during the infection-spreading phase \cite{note0}, Eq. (\ref{dd}) can be derived from mathematical models of infectious diseases. A derivation from the susceptible-infected-recovered model is found in \cite{JPSJ2021}.

Assumption (ii) for the sequential order of countermeasures can be implemented by introducing a cost function. Hereafter, we refer to the social cost per unit of time as the ``intervention cost'' in the form of $C(R_{\mbox{t}})$. The following are assumed in the $C(R_{\mbox{t}})$ function:
The condition without intervention measures corresponds to $R_{\mbox{t}}=R_{\mbox{N}}$, in which $C(R_{\mbox{N}}) = 0$. The cost should increase as the effective reproduction number decreases. 
The rate of increase in $C(R_{\mbox{t}})$ should also increase as the effective reproduction number is reduced by the countermeasures because they are adopted in sequential order of cost-effectiveness.
Thus, the following conditions can be set for the intervention cost function per unit of time, $C(R_{\mbox{t}})$ ($ 0 < R_{\mbox{t}} \le R_{\mbox{N}}$):
\begin{equation}
C(R_{\mbox{N}}) =0, 
\label{naturalcost}
\end{equation}
\begin{equation}
C(R_t - \epsilon) \ge C(R_t) \,  \mbox{ for } \forall \, \epsilon > 0,
\end{equation}
\begin{equation}
C(R_{\mbox{t}}) \, \mbox{is convex downward}.
\label{ddc}
\end{equation}

Assumption (iii) for the medical cost per unit of time yields
\begin{equation}
\frac{d M(I)}{d I} \ge 0,
\label{mc}
\end{equation}
where $M(I)$ is the medical cost. Note that the above inequality includes the case where medical cost $M(I)$ is nonlinear with respect to $I$, which contrasts with the model where medical cost is proportional to the infected population \cite{R&M}.

The total cost per unit of time is the sum of intervention and medical costs, that is, $ C(R_{\mbox{t}} (t)) + M(I(t))$.
Thus, the total cost for a given period ($t^{\prime}$, $t^{\prime \prime}$) is defined as 
\begin{equation}
\int_{t^{\prime}}^{t^{\prime \prime}} [C(R_{\mbox{t}} (t)) + M(I(t))] dt.
\label{total}
\end{equation}

With this formulation, our task is to find which countermeasure policy, ``Step by step and intensity" or ``Early and moderate countermeasure," minimizes the average of the total cost $\langle C(R_{\mbox{t}} (t)) + M(I(t)) \rangle$ 
 over a given period ($t^{\prime}$, $t^{\prime \prime}$)  (see Fig. 1).
To determine the optimized process specified by a protocol of the effective reproduction number, $R(t)$, for a given period, we must describe how the infected population increases. Thus, we must solve Eq. (\ref{dd}).
 
The solution of Eq. (\ref{dd}) at time $t$ is 
\begin{equation}
I(t) = I_0 e^{\gamma t \Delta} ,
\label{T}
\end{equation}
where $\Delta (t)$ is assumed a constant, $\Delta$, and $I_0$ is an infected population at $t=0$.
As $\Delta = R_{\mbox{t}} -1$, the intervention cost is treated hereafter as a function of $\Delta$: $C(\Delta(t))$.
Path B in Fig. 1 is set at a monotonic exponential growth rate, $\Delta (t) = \Delta^{\ast}$.
At $t=T$, it yields
\begin{equation}
I(T) = I_0 e^{\gamma T \Delta^{\ast}} .
\label{2a}
\end{equation}

For the path A$_{1}$ $( 0 \le t < T_1 )$, which is the 1st segment of path A in Fig. 1, we set 
\begin{equation}
\Delta (t) = \Delta_{\mbox{\small A}_1} =\Delta^{\ast} + \delta_1
\end{equation} 
and for path A$_2$ $( T_1 \le t < T )$, which is the 2nd segment of path A,
\begin{equation}
\Delta (t) = \Delta_{\mbox{\small A}_2} = \Delta^{\ast} - \delta_2 .
\end{equation} 

    \begin{figure}[h]
    \includegraphics[scale=1]{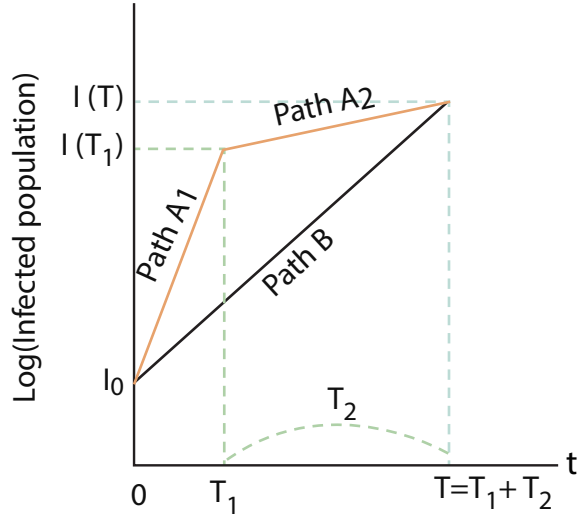} 
    \caption{Tracing infected population during the cyclic process of pandemic control. The analytic solution shows that both economic and medical costs of path A are higher than those of path B.}
    \end{figure}

Then, 
\begin{equation}
I(T_1) = I_0 e^{\gamma T_1 (\Delta^{\ast} + \delta_1)} \,  \mbox{ and}
\end{equation}
\begin{equation}
 I (T_1 + T_2) = I(T_1) e^{\gamma T_2 (\Delta^{\ast} - \delta_2)} .
\end{equation}
By using $T_1 + T_2 = T$ and Eq.(\ref{2a}), we obtain the equality
\begin{equation}
\delta_1  T_1 = \delta_2   T_2 .
\label{equality}
\end{equation}

Now, the costs between the two processes, paths A and B, can be compared. The intervention costs during the period between $t=0$ and $t=T$ are written as
\begin{equation}
\begin{split}
\int_{\mbox{path A}} C(\Delta(t)) dt  & = C(\Delta^{\ast} + \delta_1)  \, T_1 +  C(\Delta^{\ast} - \delta_2) \, T_2 ,
 \end{split}
\label{eee}
\end{equation}
\begin{equation}
\int_{\mbox{path B}} C(\Delta(t)) dt  = C(\Delta^{\ast}) T =  C(\Delta^{\ast}) T_{1} +  C(\Delta^{\ast}) T_{2}.
\end{equation}
    \begin{figure}[h]
    \includegraphics[scale=1]{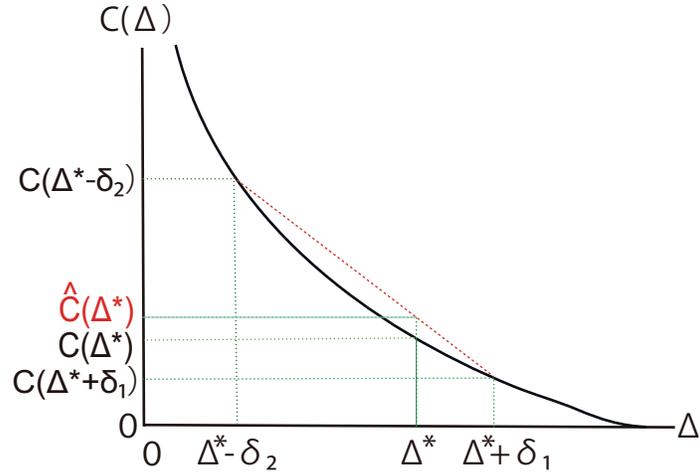} 
    \caption{An example of the functional form of $C(\Delta)$. This function need not be differentiable; it need only be a continuous function that is convex downward. The red dashed line connecting points ($\Delta^{\ast} + \delta_1$, $C(\Delta^{\ast} + \delta_1)$) and  ($\Delta^{\ast} - \delta_2$, $C(\Delta^{\ast} - \delta_2)$) takes the value $\hat{C}(\Delta^{\ast})$ at $\Delta = \Delta^{\ast}$. }
\label{Fig.K}
\end{figure}
Then, 
\begin{equation}
\int_{\mbox{path A}} C(\Delta(t)) dt  - \int_{\mbox{path B}} C(\Delta(t)) dt \\
= [C(\Delta^{\ast} + \delta_1)  - C(\Delta^{\ast}) ]  T_{1} +   [ C(\Delta^{\ast} - \delta_2) - C(\Delta^{\ast}) ]  T_2 .
\label{RHS}
\end{equation}

The convexity of $C(\Delta)$ (Eq. \ref{ddc}) leads to the inequality (see Fig. \ref{Fig.K}):
\begin{equation}
C(\Delta^{\ast}) \le \hat{C}(\Delta^{\ast}) =  \frac{\delta_1 C(\Delta^{\ast} - \delta_2) + \delta_2 C(\Delta^{\ast} + \delta_1)}{\delta_1 + \delta_2} .
\label{convex}
\end{equation}
By applying this inequality together with the equality (Eq. (\ref{equality})) into Eq. (\ref{RHS}) leads to
\begin{equation}
\begin{split}
\int_{\mbox{path A}} C(\Delta(t)) dt  - \int_{\mbox{path B}} C(\Delta(t)) dt 
 & \ge 0.
\end{split}
\end{equation} 
We therefore obtain the general inequality,  
\begin{equation}
\int_{\mbox{path A}} C(\Delta(t)) dt  \ge \int_{\mbox{path B}} C(\Delta(t)) dt .
\label{ine}
\end{equation} 
As the infected population of path A is not less than that of path B, through Eq.(\ref{mc}), the same inequality also holds for medical costs:
\begin{equation}
\int_{\mbox{path A}} M(I(t)) dt  \ge \int_{\mbox{path B}} M(I(t)) dt .
\label{me}
\end{equation} 
 Therefore, the total cost of path A is not less than that of path B:
\begin{equation}
\mbox{Total cost of path A} \ge \mbox{Total cost of path B}.
\label{ll}
\end{equation}

 The inequality has generality in the sense that it is independent of specific parameters, $\delta_{1}$, and $\delta_{2}$. Furthermore, because any integrable function can be decomposed into a step function with arbitrary precision, Eq.(\ref{ll}) holds for any process of integrable $R(t)$ on the condition that $I_{\mbox{path A}}(t) \ge I_{\mbox{path B}} (t)$ and $I(t)$ of path A reaches that of path B at $t=T$ \cite{note1}.

\begin{figure}[h]
\includegraphics[scale=1]{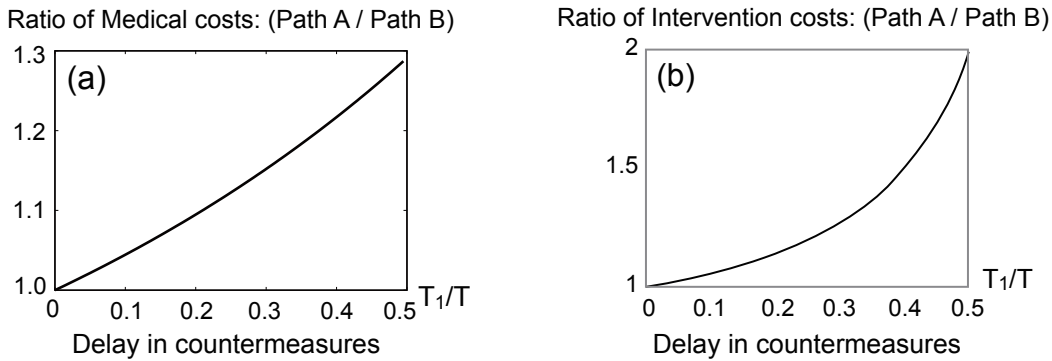} 
\caption{Example of delay effects in countermeasures in path A of Fig. 1. No measure is assumed in the 1st segment between $t=0$ and $t=T_1$. As path B is a reference and independent of the delay, the ratio of the cost of path A to that of path B shows how the delay affects costs. The greater the parameter $T_1$, the larger the delay in countermeasures. The following parameters are used for demonstration: $R_{\mbox{\footnotesize N}}=3$, $C(R_{\mbox{t}}) = (3-R_{\mbox{t}})/(R_{\mbox{t}})$,
$\Delta^{\ast} =1 \, (R_{\mbox{t}}^{\ast} =2),  \, \mbox{and} \, \gamma T = \delta_1=1$, 
 while medical cost is assumed to be proportional to the infected population, for simplicity.
(a) Medical cost, (b) Intervention cost.}
\label{ex}
\end{figure}

The results show that a society delaying countermeasures must incur more intervention and medical costs during the process. It is illustrated in Fig. \ref{ex} with the model $C(\Delta) = (\Delta_{\mbox{\footnotesize N}}-\Delta)/(1+\Delta)$. Typical examples are the countries that adopted ad-hoc countermeasures only when their medical care systems were on the verge of a breakdown. Such a policy against the pandemic is found to be non-economic based on the present theory. Notably, the present formulation and the result include the preliminary ones \cite{JPSJ2021} as a special (cyclic) case where $I(T)=I_0$.

\begin{figure}[h]
\includegraphics[scale=1]{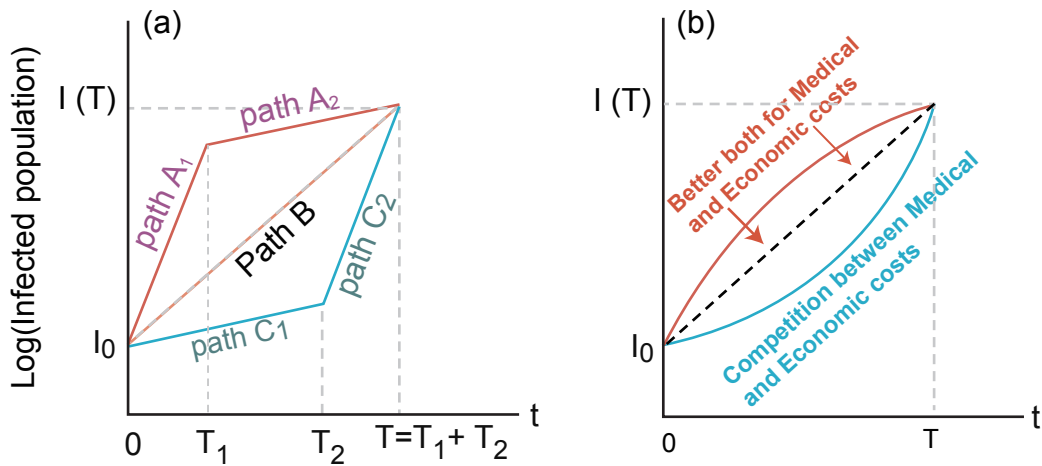} 
\caption{(a) The total cost of path A is more than that of paths B and C, although the present theory cannot solve which total cost between paths B and C is better. Here, the period and effective reproduction number of paths A$_1$ and C$_2$, as well as those of A$_2$ and path C$_1$, are identical. (b) Outline of the paper: In the upper left region: The total cost decreases as the path approaches the optimal one, represented by the broken line. The bottom right region: an optimal path cannot be found based on the present theory because it should be determined by the competition between medical and economic costs.}
\label{com}
\end{figure}

Finally, the case that the order of countermeasure is the reverse of path A of Fig. 1 is discussed to clarify this study's position. In Fig. \ref{com}(a), intense countermeasures are adopted first (path C$_1$), followed by weak countermeasures (path C$_2$), in which the effective reproduction number and the period of paths A$_1$ and C$_2$, as well as those of paths A$_2$ and path C$_1$, are identical. As the intervention cost is a function only of the effective reproduction number, the average intervention cost is the same for paths A and C. However, the average medical cost is larger in path A than in path C as the infected population of path A is always greater than or equal to that of path C. Therefore, the total cost of path C is less than that of path A. This example clarifies the structure of the finding that social delay measures have a greater total cost.

Next, let us compare the costs between paths C and B. Intervention costs of path C are the same as that of path A, as these are functions only of the effective reproduction number. 
As the intervention cost of path A is always more than that of path B, the intervention cost of path C is also more than that of path B. By contrast, the medical cost of path C is less than that of path B because the infected population of path C is less than that of path B. Thus, whether implementing a more intense countermeasure at the beginning is more beneficial than the exponentially constant countermeasure during the whole period cannot be determined by the present theory. The total cost in the bottom right region of Fig.\ref{com} can only be analyzed with a more detailed knowledge of the cost parameters. In other words, the weighting factors between medical and economic costs determine which is better in this region. For example, the more toxic the pandemic, the more the weight of the medical cost, indicating that not just early but intense countermeasures are important for reducing total cost.

This study analytically demonstrated the fundamental structure of irreversibility in the pandemic control process. Delaying measures against the spread of infection generally results in cost increases. Once the infected population increases more than the reference path B, the system is economically irreversible because extra expenditures compared with the reference path must be incurred. The results are not restricted to the specific model and are applicable as long as the infection-spreading phase is expected to last longer than the time scale of variation in the infected population. The theoretical methodology described in this paper leading to the exact solution and the generality of the findings is a novel contribution to the econophysics literature.

These general and robust findings contradict the ad-hoc policy of adopting serious measures only just before a crisis. Assuming we can estimate the maximum number of infected people that a medical care system can reasonably manage and the time required for changes in the environment surrounding the pandemic (e.g., vaccines, seasonal changes), it is important to calculate and maintain a target effective reproduction number to be implemented during that period to avoid economic irreversibility. The preliminary result \cite{JPSJ2021} is included in this study as a special case, and the present result applies not only to the COVID-19 pandemic, whether or not ``herd immunity'' exists, but also to any likely future pandemic \cite{To,Tillett}. 
However, the present theory cannot solve which countermeasure process is better if the process is in the bottom right region of Fig. \ref{com}(b). To solve this question, more detailed cost parameters are needed.

The generality established in this study is similar to that in thermodynamics \cite{Callen}, and is in contrast to the conventional methodology of mathematical economics. Thermodynamics provides a quantitative relationship among physical variables and postulates thermodynamic irreversibility, which is similar to the present result that an excess infected population is economically irreversible. The irreversibility arises from two mathematical structures: 1) the concave nature of the functional form, $C(R_{\mbox{t}})$, of the intervention cost (Eq. (\ref{ddc})), and 2) the exponential dynamics of the infected population (Eq.(\ref{T})), suggesting a similar irreversibility structure outside of econophysical phenomena.
	
As noted in the introduction, our findings were derived by assuming minimal factors concerning pandemic regulation and economics. A real system is affected by many more factors such as vaccination effect, seasonal effects, and improved ventilation. However, all the factors in this study are considered common in many real-world complex systems. 
Thus, the results are expected to be useful both as a basis for considering more realistic situations and for further analytic theory, given the importance of exact solutions to the fundamental theory in physics.

The author wishes to acknowledge K. Hirata and T. Kawakatsu for fruitful discussions and M. Arikawa and S. Takagi for critical reading of the manuscript. This work was supported by the Japanese Grants-in-Aid JSPS N. 20H00002.

\bibliography{basename of .bib file}

\begin{thebibliography}{99}
\bibitem{CEPR}
For example, articles in {\it Covid Economics, Centre for Economic Policy Research}, (Vetted and Real-Time Papers, 2020) (https://cepr.org/content/covid-economics-vetted-and-real-time-papers-0).

\bibitem{R&M}
R. Rowthorn and J. A. Maciejowski, Cost--benefit analysis of the COVID-19 disease, Oxford Rev. Econ. Pol. {\bf 36}, 38 (2020) (https://doi.org/10.1093/oxrep/graa030).

\bibitem{ERT}
M. S. Eichenbaum, S. Rebelo, and M. Trabandt, The macroeconomics of epidemics, NBER Working Paper No. 26882 (2020) (\mbox{https://www.nber.org/papers/w26882}).

\bibitem{Acemoglu}
D. Acemoglu, V. Chernozhokov, I. Werning, and M.D. Whinston, Multi-risk SIR model with optimally targeted lockdown, NBER Working Paper No. 27102 (2020) (https://www.nber.org/papers/w27102).

\bibitem{Saltelli}
A. Saltelli, G. Bammer, I Bruno {\it et al.},
Five ways to ensure that models serve society: a manifesto, Nature {\bf 582}, 482 (2020).


\bibitem{JPSJ2021}
Tsuyoshi Hondou, Economic irreversibility in pandemic control processes: Rigorous modeling of delayed countermeasures and consequential cost increases, J. Phys. Soc. Jpn. {\bf 90}, 114007 (2021).

\bibitem{Ayres}
An early discussion of the thermodynamic perspective on irreversibility in economics can be found at:
R.U. Ayres and I. Nair, Thermodynamics and economics,
Phys. Today {\bf 37}, 62 (1984).

\bibitem{note0}
During the outbreak phase, the infected population varies exponentially according to the effective reproduction number. See also Figure A$\cdot$1 in \cite{JPSJ2021}. 

\bibitem{BF}
F. Brauer, in {\it Mathematical Epidemiology}, edited by F. Brauer, P. Driessche, and J. Wu (Springer-Verlag, Berlin, Heidelberg, 2008), Chap. 2 (https://www.springer.com/gp/book/9783540789109).

\bibitem{Boardman}
A. E. Boardman, D. H. Greenberg, A. R. Vining, and D. L. Weimer, {\it Cost-Benefit Analysis: Concept and Practice} (Prentice Hall, Upper Saddle River, NJ, 2001), 2nd Ed.
 (\mbox{https://www.cambridge.org/highereducation/books/costbenefit-analysis/484720E57798B7E7A29C7156407CD4A1}).

\bibitem{Nas}
T. F. Nas, {\it Cost-Benefit Analysis: Theory and Application} (Lexington Books, Maryland, MD, 1996) (https://rowman.com/ISBN/9781498522526/Cost-Benefit-Analysis-Theory-and-Application-2nd-Edition).


\bibitem{note1}
As any path can be decomposed into pairs of small segments, $A_{1i}$ and $A_{2i}$ that corresponds to a segment $B_{i}$ of path B, Eq.($\ref{ine}$) holds for a curved line.
 
\bibitem{To}
K. K. W. To {\it et al.}, Coronavirus disease 2019 (COVID-19) re-infection by a phylogenetically distinct severe acute respiratory syndrome coronavirus 2 strain confirmed by whole genome sequencing, ciaa1275, Clin. Infect. Dis. (2020) (https://academic.oup.com/cid/advance-article/doi/10.1093/cid/ciaa1275/5897019).

\bibitem{Tillett}
R. Tillett {\it et al.}, Genomic evidence for reinfection with SARS-CoV-2: a case study, Lancet {\bf 21}, 52 (2020) (https://www.thelancet.com/journals/laninf/article/PIIS1473-3099(20)30764-7/fulltext.

\bibitem{Callen}
H. Callen, {\it Thermodynamics} (John Wiley \& Sons, Hoboken, NJ, 1960) 


\end{thebibliography}

\end{document}